\documentclass[conference]{IEEEtran}
\ifCLASSINFOpdf
\else
\fi
\usepackage{tikz}
\usepackage[sumlimits,intlimits]{amsmath}
\usepackage{amsmath,amsfonts,amssymb}%
\usepackage{bm}%
\usepackage{graphicx,graphics}%
\usepackage{cases}%
\usepackage[noadjust]{cite}%
\usepackage{color}%

\usepackage{verbatim}
\usepackage{enumerate}
\usepackage{algorithm}
\usepackage{balance}
\usepackage{multirow}
\usepackage{stfloats}
\usepackage{blindtext}
\usepackage{xspace}
\usepackage{booktabs}
\usepackage{subfig}
\usepackage{cite}
\usepackage[justification=centering]{caption}
\begin{document}
\title{Waveguide-Fed Lens Based Beam-Steering Antenna For 5G Wireless Communications   }


\author{\IEEEauthorblockN{Saeideh Shad, Shafaq Kausar, Hani Mehrpouyan}
	\IEEEauthorblockA{Department of Electrical and \ Computer Engineering\
		Boise State Univeristy\\
		Email: saeidehshad@boisestate.edu, shafaqKausar@boisestate.edu, hanimehrpouyan@boisestate.edu  }
}
%Telephone: (800) 555--1212, Fax: (888) 555--1212}


\maketitle

\vspace*{-00pt}
\begin{abstract}
	
	\vspace*{-0pt}
	In this paper, a two-dimensional cylindrical Lens antenna based on the parallel plate technique is designed. It supports beam-steering capability of $58^0$ at 28 GHz. The antenna is composed of low loss rectangular waveguide antennas, which are positioned around a homogeneous cylindrical Teflon lens in the air region of two conducting parallel plates. The Beam scanning can be achieved by switching between the antenna elements. The main advantages of our design include its relative simplicity, ease of fabrication, and high-power handling capability. Compared to previous works including a curvature optimization for the plate separation of the parallel plates, the proposed antenna has a constant distance between plates.  At the 28 GHz, the maximum simulated gain value is about 19 dB. Furthermore, the designed antenna only deviates about 0.4 dB over the $58^0$ scan range.

\end{abstract}


%no keywords

% \fi
\IEEEpeerreviewmaketitle
\vspace*{-0.2cm}
\section{Introduction}

 Millimeter-wave antenna design is considered as the first step for realizing mm-wave wireless communication systems. Design requirements for such antennas include highly directional patterns. Based on this demand, Luneburg lens (LL) antenna is an attractive choice at next generation wireless communications (5G) systems to create high gain directional radiation patterns \cite{citation1 , citation2}. Recently, several works of two-dimensional parallel plates waveguide (PPW) designs with fan beam scanning capability have been a subject of extensive research \cite{citation4,citation5}. In this letter, a simple structure of PPW inspired multibeam antenna is demonstrated. In contrast to previous works used planar microstrip feeds, we are using metallic waveguides which have low loss, compact and slim features to fit between plates. Furthermore, the two parallel plates are separated by a constant distance. However, in previous PPW antennas the distance between the two parallel plates varies along with the plate’s length, forming a non-linear curvature.  
 
\vspace*{-0cm}
	\begin{figure}[t]
		\includegraphics[width=85mm]{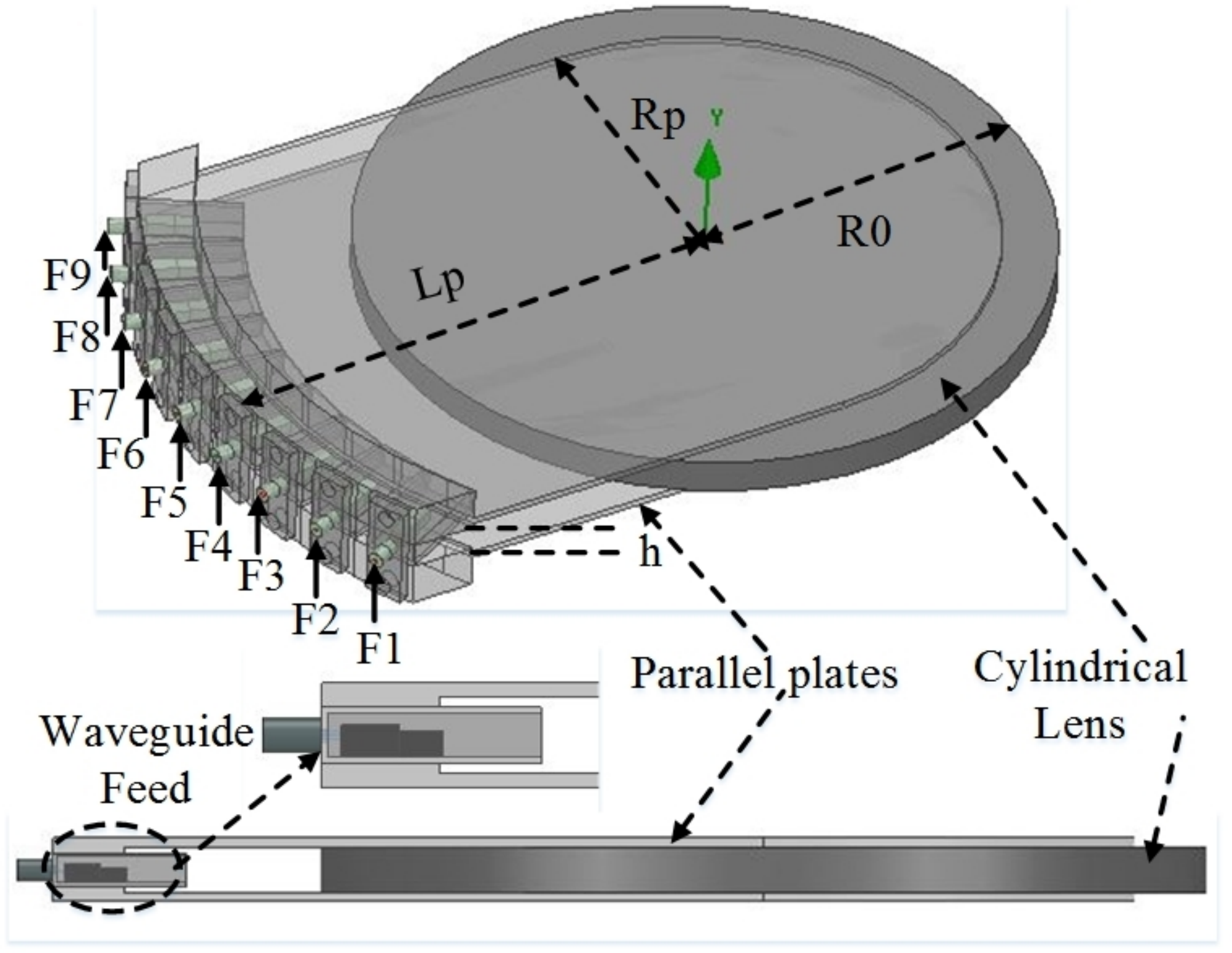}
		\caption{Geometry of the waveguide-fed cylindrical dielectric lens (${R_o}/{\lambda_0}$ = 4.6, ${R_p}/{\lambda_0}$ = 3.7, ${h}/{\lambda_0}$ = 0.54).}
		 \label{fig:view}
	\vspace{0cm} 
	\end{figure}

\vspace{-0.2cm} 
\section{Design and Configuration}

Fig. 1 shows the three-dimensional view of the proposed
beamsteering antenna. It mainly consists of three parts:
feeding-network, the dielectric lens and conductive two parallel plates. The proposed lens with relative dielectric constant of $\epsilon_r = 2.1$ and $tan\delta = 0.0002$ has cylindrical cross section sandwiched between the plates. To estimate lens parameters, from antenna theory \cite{citation3}, the E-plane half-power beamwidths of the LL is given by the expression:
\begin{equation}
BW_E\approx 29.4(\frac{\lambda_0}{R_0})
\vspace{-0.2cm}
\end{equation}

where $\lambda_0$ represents the free space wavelength and $R_0$ is the radius of cylindrical lens. A radius of $\approx$ 49.2 mm is required to produce a radiation pattern ($BW_E$) of $6.5^0$ for the operating frequency of 28 GHz. A simple coaxial-line to rectangular-waveguide (RW) transition has been designed to feed antenna. The transition consists of a stepped impedance and mode transformer in waveguide structure. A standard 2.92mm-type connector has been used as coaxial connector. For the PPW, the plate separation is considered in the range of ${\lambda_0}/{2}<$h$<{\lambda_0}$ to excite $TE_{10}$ mode between two plates. Waveguide feed is embedded in the initial section of the parallel plates. The phase center of the RW feed needs to be coincident with the focal point of the lens. The beam steering capability is achieved by arranging of nine RW elements in arc direction in azimuthal plane with respect to dielectric lens. Then, feeds are placed in a focal arc with $7.2^0$ spacing. Each feeding element is represented with F1, F2,...F9 [Fig. 1].

\section{Simulation Results}
\begin{figure}[]
  	\includegraphics[width=\linewidth]{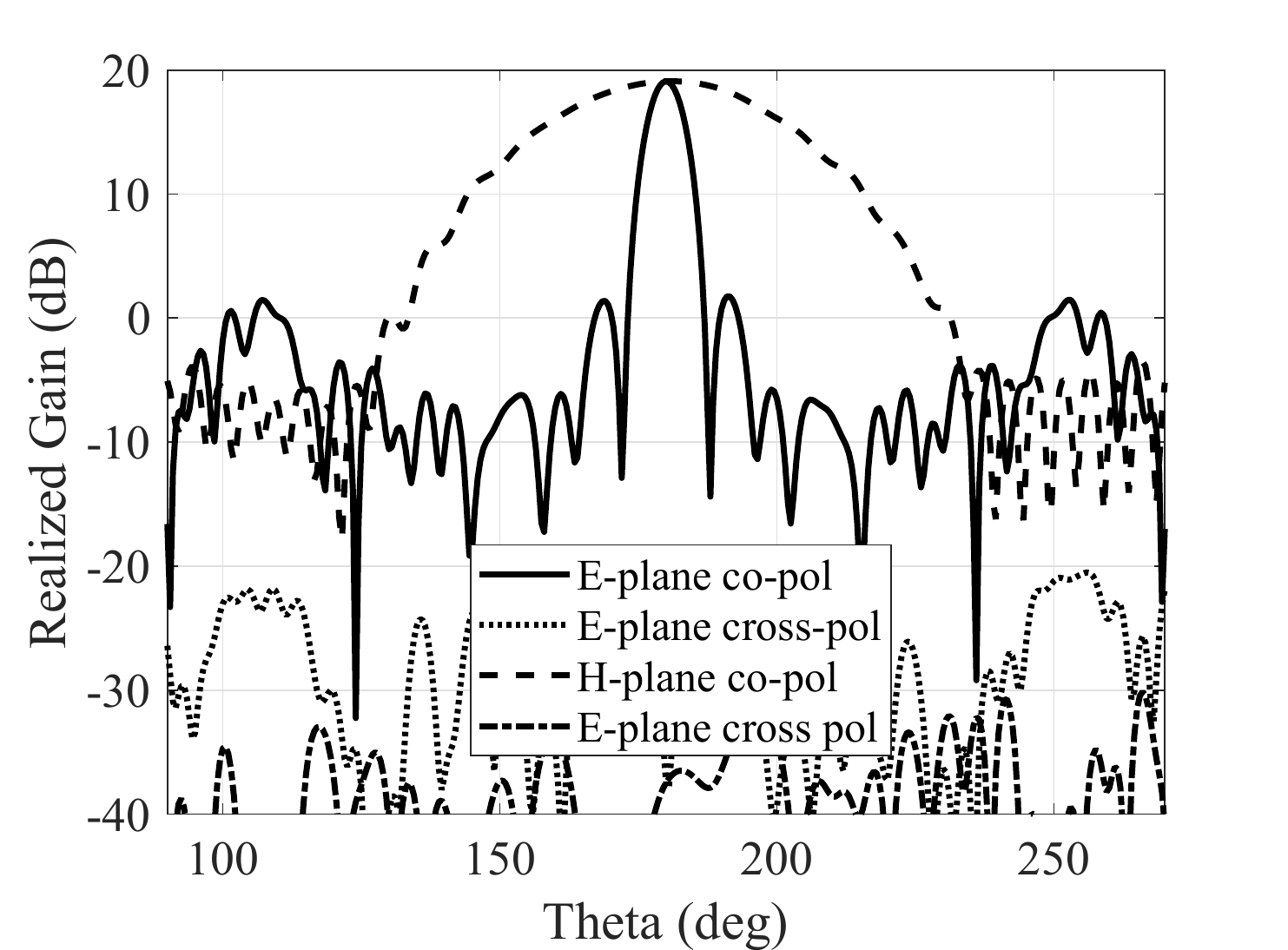}
  	\caption{The simulated radiation pattern at 28 GHz of the feeding element.}
  	\label{fig:Gainonefeed}
	
  \end{figure}

\begin{figure}[]
  
  	\includegraphics[width=\linewidth]{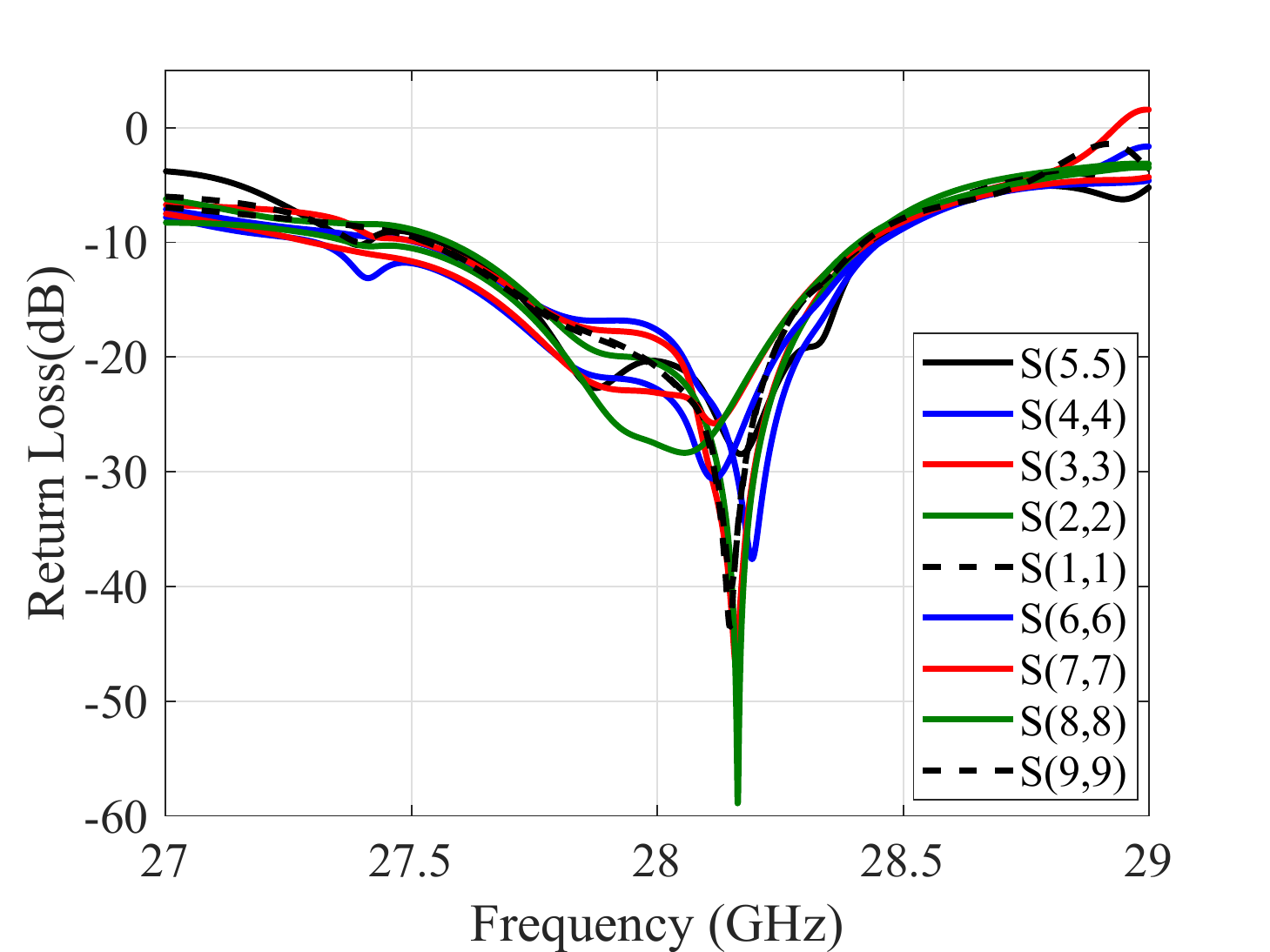}
  	\caption{The simulated reflection coefficient of the Lens based beam-steering antenna at 28 GHz excited by ports: $F_1-F_9$.}
  	\label{fig:S11allport-2}
  	\vspace*{-0.3cm}
  \end{figure}

\begin{figure}
  	\includegraphics[width=\linewidth]{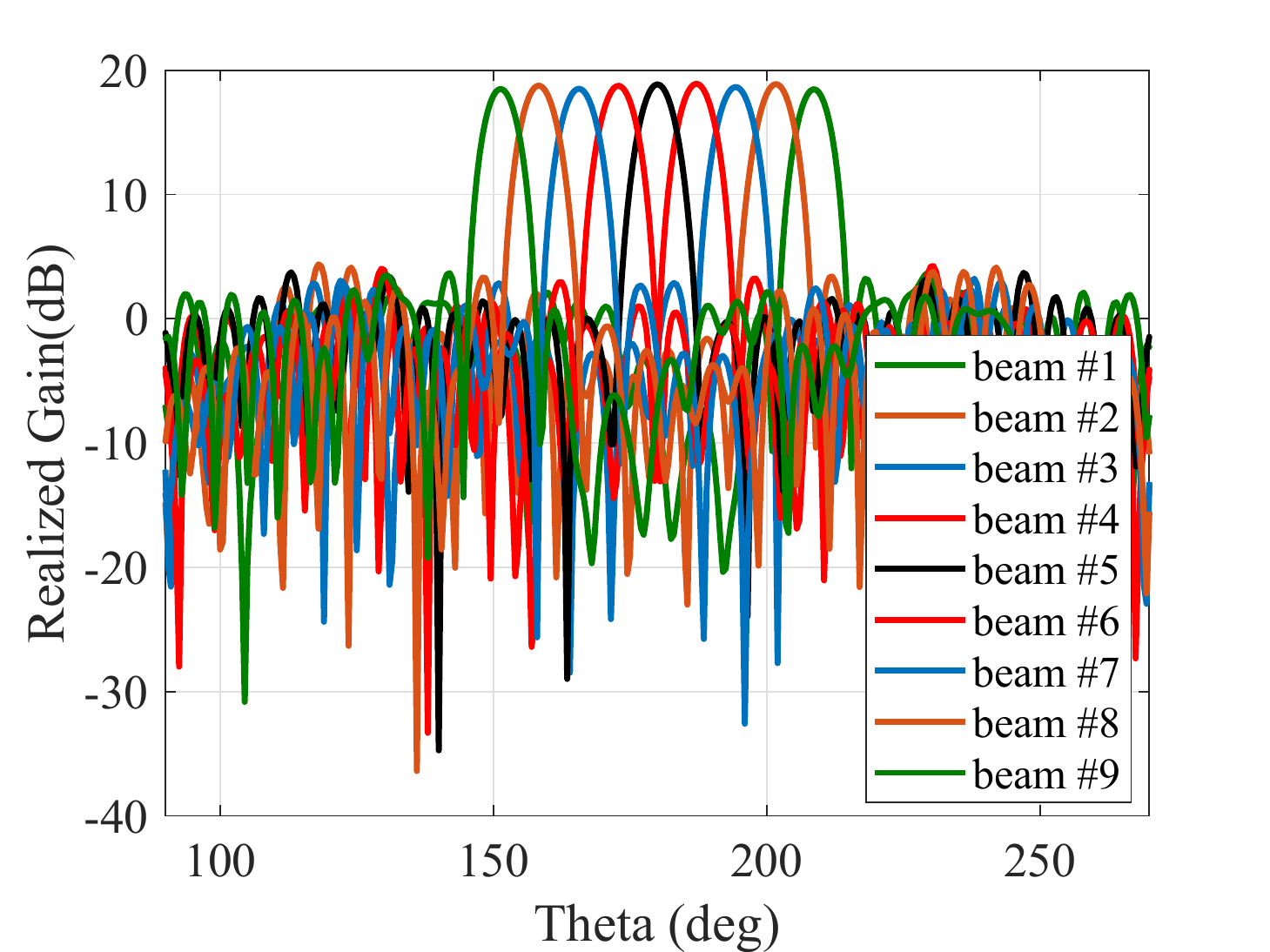}
  	\caption{The E-plane radiation patterns of the Lens based beam-steering antenna at 28 GHz.}
  	\label{fig:Total}
  \end{figure}

The single RW feed with a coaxial transition has been designed and simulated at 28 GHz.  Initially, we started with one feeding element to illuminate dielectric lens. Since it is desirable to have good illumination over an extended portion of the cylindrical lens, positioning of the feed is a critical part of the simulation. According to \cite{citation6}, first we placed the RW feed at a 0.4${Ro}$ distance from the edge of the lens. From this approximation, for achieving maximum gain and less sidelobe level the aperture of the RW was swept in a distance from the lens surface to determine the optimal feed position. Ultimately, the optimal position is achieved at 0.32${Ro}$ distance from the edge of the lens. The E-palne and H-plane radiation patterns of the feeding element integrated with parallel plate and lens is shown in Fig. \ref{fig:Gainonefeed}. At the Plate spacing of 0.54${\lambda_0}$, the simulated 3-dB beamwidth in E-plane and H-plane is about $6.4^0$ and $40^0$ degree respectively. Since the cylindrical lens has a continuous focal arc around its circumference, multiple feed elements placed next to each other with a angular spacing of 7.2 degree. Fig. \ref{fig:S11allport-2} depicts the simulated reflection coefficient of the multiple RW feeds versus frequency (GHz). It can be seen that the simulated reflection coefficient is less than -18.0 dB at 28 GHz for all ports. Due to symmetry around the center port, symmetrical ports are shown with the same color. Ideally, signals of two adjacent ports will interfere with each other. By exciting each port, a distinct beam is created in the desired direction. The radiation pattern of the resulting beam-steering for all feeds is shown in Fig. \ref{fig:Total}. Table I demonstrates the radiation characteristics achieved by each excited port. As displayed, multiple beams within a range of $58^0$ with a gain variation of less than 0.4 dB resulted in a 3-dB beamwidth of about $6.15^0${--} $6.42^0$. 
\vspace{-0.2cm}  
\begin{table*}[]
	\renewcommand{\arraystretch}{1.2}
	
	\caption{Radiation characteristics of the nine individual ports of the Lens based beam-steering antenna }
	\label{tab2}
	\centering
	\begin{tabular}{cccccclllclc}
		
		\hline\hline
		
		port & 1 & 2 & 3 & 4 & 5 & 6 & 7 & 8 & 9 \\ 
		\hline
		Beam width(deg) & 6.37 & 6.15 & 6.56 & 6.22 & 6.38 & 6.36 & 6.42 & 6.15 & 6.37 \\
		Peak gain (dB) & 18.5 & 18.9 & 18.7 & 18.9 & 18.8 & 18.7 & 18.5 & 18.9 & 18.5  \\
		Beam direction (deg) & 151 & 158.5 & 166 & 173 & 180 & 187 & 194 & 201.5 & 208.5 \\
		\hline	
	\end{tabular} 
\vspace*{-0.6cm}                   
\end{table*}

\vspace{-0cm} 
\section{Conclusion}
\vspace{0cm}
 A simple and low loss design of PPW lens based antenna with beam steering capability has been designed at 28GHz. The antenna is fed with an array of metallic rectangular waveguides to overcome the transmission losses of conventional PPW antennas at high frequencies. The simulated results show a good impedance bandwidth and good radiation patterns at the operation frequency.





%\bibliographystyle{IEEEtran}

\end{document}